\documentclass[twocolumn]{aa}
\usepackage{graphicx}
\usepackage{txfonts}

\begin{document}

   \title{Gravitational Wave Driven Mergers and Coalescence
Time of Supermassive Black Holes}


   \author{Fazeel Mahmood Khan
          \inst{1}
          \and
          Peter Berczik\inst{2,3}
          \and
          Andreas Just \inst{4}
          }
   \institute{Department of Space Science, Institute of Space Technology, PO Box 2750 Islamabad, Pakistan\\
              \email{khanfazeel.ist@gmail.com}
         \and
            Main Astronomical Observatory, National Academy of Sciences of Ukraine, MAO/NASU, 27 Akad. Zabolotnoho St. 03680 Kyiv, Ukraine\\
              \email{berczik@mao.kiev.ua}
         \and
            National Astronomical Observatories and Key Laboratory of Computational Astrophysics, Chinese Academy of Sciences, NAOC/CAS, 20A Datun Rd., Chaoyang District Beijing 100012, China NAOC, Bejing
              \email{berczik@nao.cas.cn}
         \and
             Astronomisches Rechen-Institut, Zentrum f\"{u}r Astronomie der Universit\"{a}t Heidelberg, M\"{o}nchhofstr. 12--14, D-69120 Heidelberg, Germany\\
              \email{just@ari.uni-heidelberg.de}
             }

   \date{Accepted ; Astronomy \& Astrophysics}

 
  \abstract
   {}
   {The evolution of Supermassive Black Holes (SMBHs) initially embedded in the
centres of merging galaxies realised with a stellar mass function (SMF) is studied from the onset of galaxy mergers till
coalescence. Coalescence times of SMBH binaries are of great importance for black hole evolution and gravitational wave detection studies.}
   {We performed direct N-body simulations using the highly efficient
and massively parallel phi-GRAPE+GPU code capable to run on GPU supported high
performance computer clusters. Post-Newtonian terms up to order 3.5 are used
to drive the SMBH binary evolution in the relativistic regime. We performed a large set of simulations with three different slopes of the central stellar cusp and different random seeds. The impact of a SMF on the hardening rate and the coalescence time is investigated.}
   {We find that SMBH binaries coalesce well within one billion years when our models are scaled
to galaxies with a steep cusp at low redshift. Here higher central densities provide
larger supply of stars to efficiently extract energy from the SMBH binary orbit
and shrink it to the phase where gravitational wave (GW) emission becomes
dominant leading to the coalescence of the SMBHs. Mergers
of models with shallow cusps that are representative for giant elliptical galaxies having central cores
result in less efficient extraction of binary's orbital energy due to the lower stellar
densities in the centre. However, high values of eccentricity witnessed for SMBH
binaries in such galaxy mergers ensure that the GW emission dominated phase
sets in earlier at larger values of the semi-major axis. This helps to compensate for the
less efficient energy extraction during the phase dominated by stellar encounters
resulting in mergers of SMBHs in about one Gyr after the formation of the
binary. Additionally, we witness mass segregation in the merger remnant resulting in enhanced SMBH binary hardening rates. We show that at least the final phase of the merger in cuspy low mass galaxies would be observable with the GW detector eLISA.}
   {}


\maketitle

\section{Introduction}

Gravitational waves (GWs) were predicted by Einstein shortly after presenting his
General theory of Relativity (GR) \citep{ein16,ein18}. Indirect evidence that GWs exist came from studies of the orbital decay of binary pulsars in accordance with
GR \citep{hulse75}. Recently, the first direct measurement of GWs was accomplished by the observation of the merging event of the stellar mass black hole binary
GW150914 with LIGO \citep{Abbott16}. 
Meanwhile three more events including a neutron star -- neutron star merger were observed with LIGO and VIRGO  \citep{abb17}.
Observations of GWs coming from various sources at various cosmic epochs would open an entirely new window to study the universe,
currently beyond the capabilities of electromagnetic probes.
Supermassive black hole (SMBH) binaries are considered as promising sources
of GWs \citep{begelman80}. Observations of GWs emitted during the final phase of in-spiral would provide the merger history of galaxies as a function of redshift leading to important constraints on SMBH and galaxy formation and evolution
scenarios. The International Pulsar Timing Array (IPTA) is searching for GWs
from coalescing binary SMBHs in the mass range $10^7 - 10^9\,M_\sun$ up to redshift $z=2$ \citep{hobbs10,des16,rea16,nan15,ver16}.
In future the (evolved) Laser Interferometer Space Antenna (eLISA, LISA) is expected to
detect GWs to much larger redshifts ($z \sim 10$) \citep{amaro13,goat16,lisa17}.

SMBH binaries form as a result of mergers between two sizable galaxies, each
hosting a central SMBH, which are ubiquitous in galaxy cores \citep{mcc13,kor14}. Observational searches for SMBH binaries are going on with a handful cases of two well
separated accreting SMBHs seen as AGNs as well as circumstantial evidence for
bound Keplerian binaries \citep{kom03,bogda15,gra15}. In the merger remnant, the evolution of SMBHs happen in three phases, each characterised by a distinct physical process. Dynamical friction \citep{chand43,just11} is responsible for the initial sinking of SMBHs in the merger remnant bringing  the
SMBHs close enough that they form a binary system. Dynamical friction efficiently extracts energy from the binary orbit till the binary gets hard. By this point
of time, dynamical friction ceases to be efficient while SMBHs are separated by
parsec scale distances. In the second phase, stars on orbits intersecting the binary orbit
extract energy from the SMBH binary by the slingshot mechanism during 3-body encounters bringing the black holes closer. If it is efficient to bring the SMBHs in the binary close
enough (milli parsec separations), GW emission in the third and final phase drains
out the remaining energy in the binary orbit leading the SMBHs to coalesce.
How efficient the SMBH binary evolve in the 3-body scattering phase such that
the separation between the SMBHs shrinks to the GW dominated regime depends strongly
on the orbit contents of the host galaxy (merger remnant) \citep{mer04,li15,gua17}. Earlier it was shown that
for galaxy shapes close to sphericity, SMBH binaries may stay longer than a
Hubble time in the 3-body scattering regime \citep{mak04,ber05}. However strongly flattened or mildly triaxial shapes, a natural product of galaxy mergers, have shown an effective shrinking of SMBH binary's semi-major axis to the point where GW
emission dominates \citep{ber06,khan11,preto11,gua12,khan13,vasi15}. The hardening rate and eccentricity of the binaries depend strongly on the central stellar density profile and hence the
estimated coalescence times of SMBHs in binaries \citep{khan12}.
In previous studies of SMBH binary evolution in equilibrium galaxy models
or in mergers of galaxy bulges, coalescence times were obtained \citep{khan15,holley15,ses15,ran17} by extrapolating the
nearly constant hardening rate of SMBH binaries in the 3-body scattering phase to the
GW dominated regime and then using orbit averaged expressions \citep{pet63} for the hardening by GW
emission. 

Here we study the SMBH binary evolution in mergers of galaxy spheroids
having various stellar density profiles towards the centre, like in \cite{khan12}
but this time following the binary evolution into the relativistic regime for a complete set of mergers by including post-Newtonian (PN) terms up to order 3.5 in the equation of motion of the SMBH binary. 
The effect of a stellar mass function on the SMBH binary evolution is not well studied. On one hand side it is known from three body scattering investigations \citep{hills80,ses10} that for a uniform stellar population the hardening rate should be independent of the stellar mass in the low mass regime and reduced for higher intruder masses above 1:10 with respect to the secondary SMBH mass. On the other hand a mass segregated system should have an enhanced hardening rate, because the velocity dispersion of the high mass end is smaller leading to an enhanced contribution to the energy extraction.
In order to investigate this effect we
introduced a stellar mass function (SMF) for particles in the merging galaxies in order to allow for mass segregation effects.

This paper is arranged as follows: section 2 describes our models and their
scalings. It also includes numerical codes and hardware used to perform the galaxy
merger and SMBH binary evolution simulations. Section 3 presents the results of
our study. Section 4 summarises and concludes our study.

\section{Simulation Setup and Numerical Techniques} \label{ini-cond}
\subsection{Galaxy Models}\label{gal-mod}

We setup our initial galaxies by spherical isotropic distributions of stars such that the density distribution satisfies a \cite{deh93} profile

\begin{equation}
\rho(r) = \frac{(3-\gamma)M_\mathrm{gal}}{4\pi}\frac{r_{0}}{r^{\gamma}(r+r_{0})^{4-\gamma}}, \label{denr}\\
\end{equation}

Here $M_\mathrm{gal}$ denotes the mass of the galaxy, $r_0$ its scale radius, and $\gamma$ represents the slope of the inner density profile. We generated galaxy models for three different values of 
$\gamma$, $\gamma = 0.5, 1.0, 1.5$ (series A, B, C, respectively) in order to cover the observed range of density profiles of galactic nuclei. A particle having mass equal to 1 percent of the mass of the galaxy is placed at the centre to represent the SMBH. This SMBH mass ($M_\bullet$) is a few times greater compared to the observed $M_\bullet$-$M_\mathrm{gal}$ relation. Therefore our models can be viewed as a representation of the central parts of galaxies/bulges. Positions and velocities are assigned to the stars by numerically computing the distribution function in the combined potential of black hole and stars such that our models are in dynamical equilibrium.
In our ``model units'' $M_\mathrm{gal} = G = r_{0} = 1$ for the primary (more massive) galaxies.
The masses of the smaller, secondary, galaxies and their SMBHs are scaled down by a factor $q$. The smaller galaxies follow the same density profile, as the primary galaxies, but have smaller masses and different scale radii $r_\mathrm{0,s}$. The size ratio of the galaxies and the corresponding mass ratio $q$ are related as $r_\mathrm{0,s}/r_\mathrm{0} \propto \sqrt{q}$. Both primary and secondary galaxies have the same number of particles $N$. For this study we choose $q = 0.25$ and $N=200,000$. The SMBH binary evolution in the merger remnant realised with $400,000$ stellar particles is expected to be $N-$independent \citep{ber06,preto11,khan11}. The basic setup is very similar to the models in \cite{khan12} but with slightly smaller particle number and larger SMBH mass in order to speed up the simulations. 

For each $\gamma$, we generated three galaxy models with different random initialisation (seed) for both major and minor galaxies. 
In order to study the effect of a stellar mass function (SMF) and the corresponding mass segregation on the SMBH binary evolution each galaxy model was given a mass function according to a Salpeter IMF in the mass range of 0.08 to 8\,$M_\sun$
with a mean mass of 0.25\,$M_\sun$, 
also with three different random realisations (see table \ref{TableA}).
In our simulations each particle represents a number of stars with same velocity and mass. Since the evolution times of the SMBHs are of the order of a Gyr and we are mainly interested in a possible effect of mass segregation, we did not include mass loss by stellar evolution. For mass segregation the dynamic range of particle masses is the most important parameter. Our choice of a factor of 100 between most and least massive particles corresponds to the mass range of stellar black holes and the lower main sequence of an old population. In this sense the chosen mass function should be seen as a very rough representation of the mass function of a real galaxy.

In each galaxy the mass ratio of the SMBH and the most massive stellar particle is 1:62 and after the merger the maximum mass ratio of the secondary SMBH and the maximum stellar mass is 1:15.5, which is still in the limit of small intruder mass for the 3-body scattering events. In the test simulations discussed in Sect.\,\ref{N-test} the number of particles in increased by a factor of 5 leading to a maximum mass ratio of 1:77, which is sufficient for quantifying the hardening rate but still problematic for the investigation of the eccentricities due to large fluctuations.

For each $\gamma$ value (series A, B, C), we have ten runs, the $0^{th}$ run represents a galaxy merger of galaxies having equal mass stellar particles in each galaxy. The remaining nine runs are three galaxy models each having three different random realisations of the SMF (see table \ref{TableA}).

\begin{table*}
\caption{Galaxy merger runs} 
\centering
\begin{tabular}{c c | c c | c c | c c}
\hline
Run & $\gamma$ & Run & $\gamma$ & Run & $\gamma$ & Galaxy(seed) & SMF(seed)\\
\hline
A0 & 0.5 & B0 & 1.0 & C0 & 1.5 & seed1 & No SMF\\
A1 & 0.5 & B1 & 1.0 & C1 & 1.5 & seed1 & seed1\\
A2 & 0.5 & B2 & 1.0 & C2 & 1.5 & seed1 & seed2\\
A3 & 0.5 & B3 & 1.0 & C3 & 1.5 & seed1 & seed3\\
A4 & 0.5 & B4 & 1.0 & C4 & 1.5 & seed2 & seed1\\
A5 & 0.5 & B5 & 1.0 & C5 & 1.5 & seed2 & seed2\\
A6 & 0.5 & B6 & 1.0 & C6 & 1.5 & seed2 & seed3\\
A7 & 0.5 & B7 & 1.0 & C7 & 1.5 & seed3 & seed1\\
A8 & 0.5 & B8 & 1.0 & C8 & 1.5 & seed3 & seed2\\
A9 & 0.5 & B9 & 1.0 & C9 & 1.5 & seed3 & seed3\\
\hline
\end{tabular}\label{TableA}
\\
\tablefoot{Columns 1, 3, 5: Merger runs. Columns 2, 4 ,6: $\gamma$ for the galaxies. Columns 7, 8: Random seed to initialise the galaxy model and the SMF for the galaxies.}
\end{table*}

\subsection{Initial orbits and scaling to real galaxies} \label{ini-orb}

For each run we set two galaxies (primary and secondary galaxy) at apo-centre on eccentric orbits with $e = 0.75$. The initial separation between the centres of the merging galaxies in our simulations is 15 in our model units.

We choose three different galaxies M87, M31, MW for the physical scaling of our models. We use the observed mass of the SMBH and the velocity dispersion in these galaxies to calculate the sphere of influence $r_\mathrm{h}$ of the SMBH. Then we compare our model SMBH and its sphere of influence to the observed ones to get the physical scaling of our models (see table \ref{scale}). M87 represents giant elliptical galaxies, which typically have very shallow cusps or a core in the centre and thus represent our $\gamma = 0.5$ (series A) shallow cusp models. To scale our series B which have a $\gamma = 1.0$ inner density slope we choose M31 as representative. The physical parameters of the Milk Way centre are used to scale our $\gamma=1.5$ steep cusp models of series C. The details of the scaling parameters are given in table \ref{scale}.

\begin{table*}
\caption{Physical scalings for our galaxy models} 
\centering
\begin{tabular}{c c c c c c c c c}
\hline
\hline
Series & Galaxy & $M_{\bullet}$ & $\sigma_{\star}$ & $r_h$ & $TU$ & $LU$ & $MU$ & $c$\\
 &  & [$M_\sun$] &  [km\,s$^{-1}$] & [pc] & [Myr] & [kpc] & [$M_\sun$] & [$LU\,TU^{-1}$]\\
\hline
A & M87 & $6.05 \times 10^9 $& $325$ & $255$& $3.07$& $2.95$& $6.05 \times 10^{11}$& $320$\\
B & M31 & $1.63 \times 10^8$ &$169$   & $21.75$& $1.01$& $0.42$& $1.63 \times 10^{10}$& $733$\\
C & MW & $4.6 \times 10^6 $& $103$& $1.4$& $0.62$& $.092$   &$4.6 \times 10^8$& $2044$\\
\hline
\end{tabular}\label{scale}
\\
\tablefoot{Columns from left to right; (1) Initial galaxy model series, (2) Reference galaxy chosen for physical scalings of models in column (1), 
(3) Observed SMBH mass in the reference galaxy, (4) observed velocity dispersion (5) observed influence radius, (6, 7, 8) model units of time, length, and
mass, respectively, and (9) value of $c$ (speed of light) that we use in our simulations in model units. 
}
\end{table*}

\subsection{Numerical code} \label{num-code}

The numerical simulations of the galaxy mergers are performed using an updated version of the direct $N$-body code $\phi-GRAPE$ originally designed to run on GRAPE cards. Our updated code ($\phi$-GRAPE+GPU) is capable of running on Graphic Processing
Units (GPUs) supported massively parallel clusters. For the pairwise force calculations 

we use a softening parameter equal to $10^{-5}$ in model units for the stars and no softening for the SMBHs. For the pairwise forces we apply the rms values of the softening leading to a softening of $7\times10^{-6}$ for star-SMBH interactions and no softening for the SMBH-SMBH interaction.

Relativistic effects are taken into account by incorporating post-Newtonian terms up to order PN3.5 in the SMBH binary's equation of motion \citep{bla06}. More details on the simulation code can be found in \citet{khan13}. We used the Laohu cluster of the National Astronomical Observatories of Chines Academy of Science to perform our simulations. 

\section{SMBH Binary Evolution}\label{evol}

\begin{figure}
\centerline{
  \resizebox{0.9\hsize}{!}{\includegraphics[angle=270]{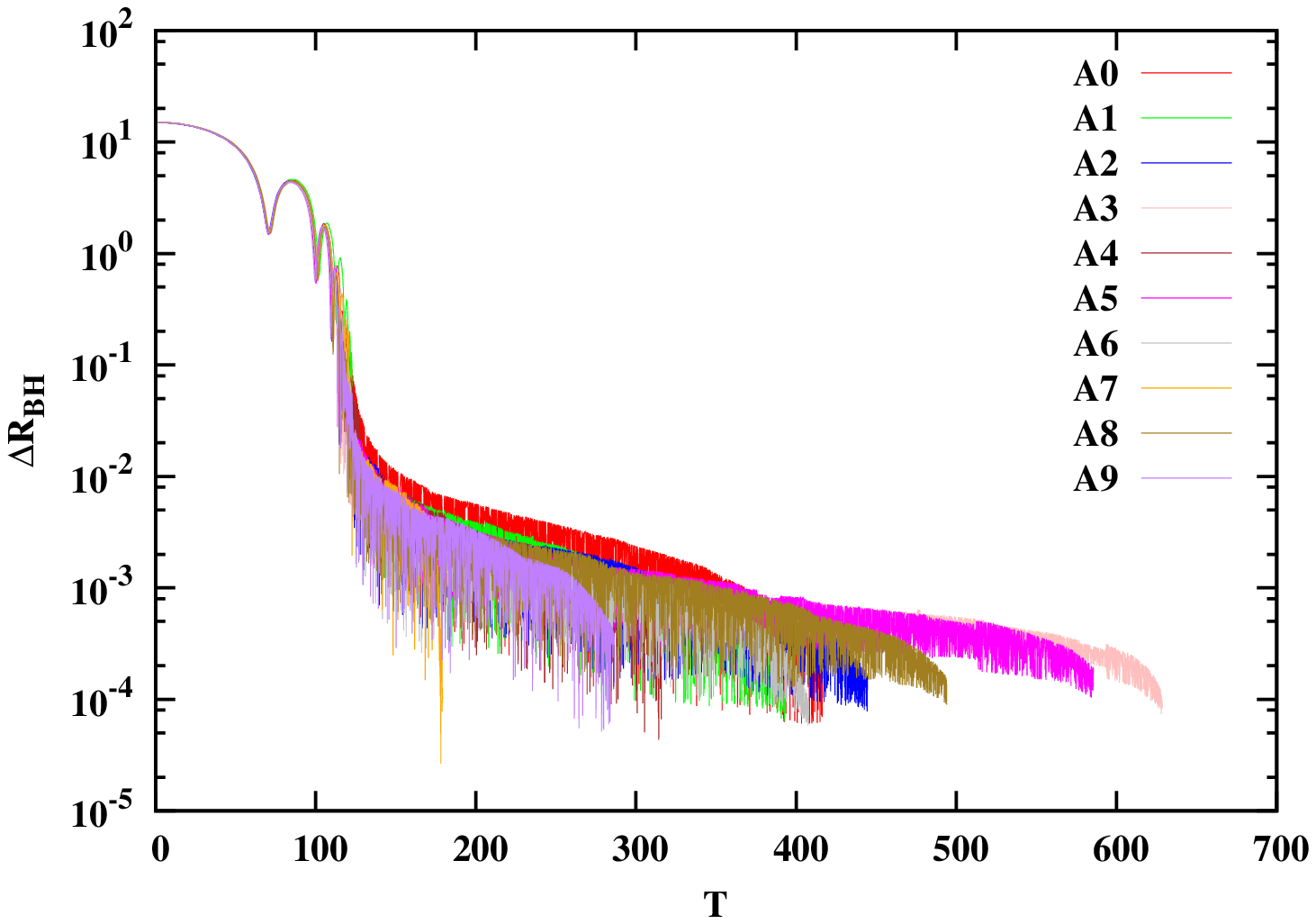}}
  }
\centerline{
  \resizebox{0.9\hsize}{!}{\includegraphics[angle=270]{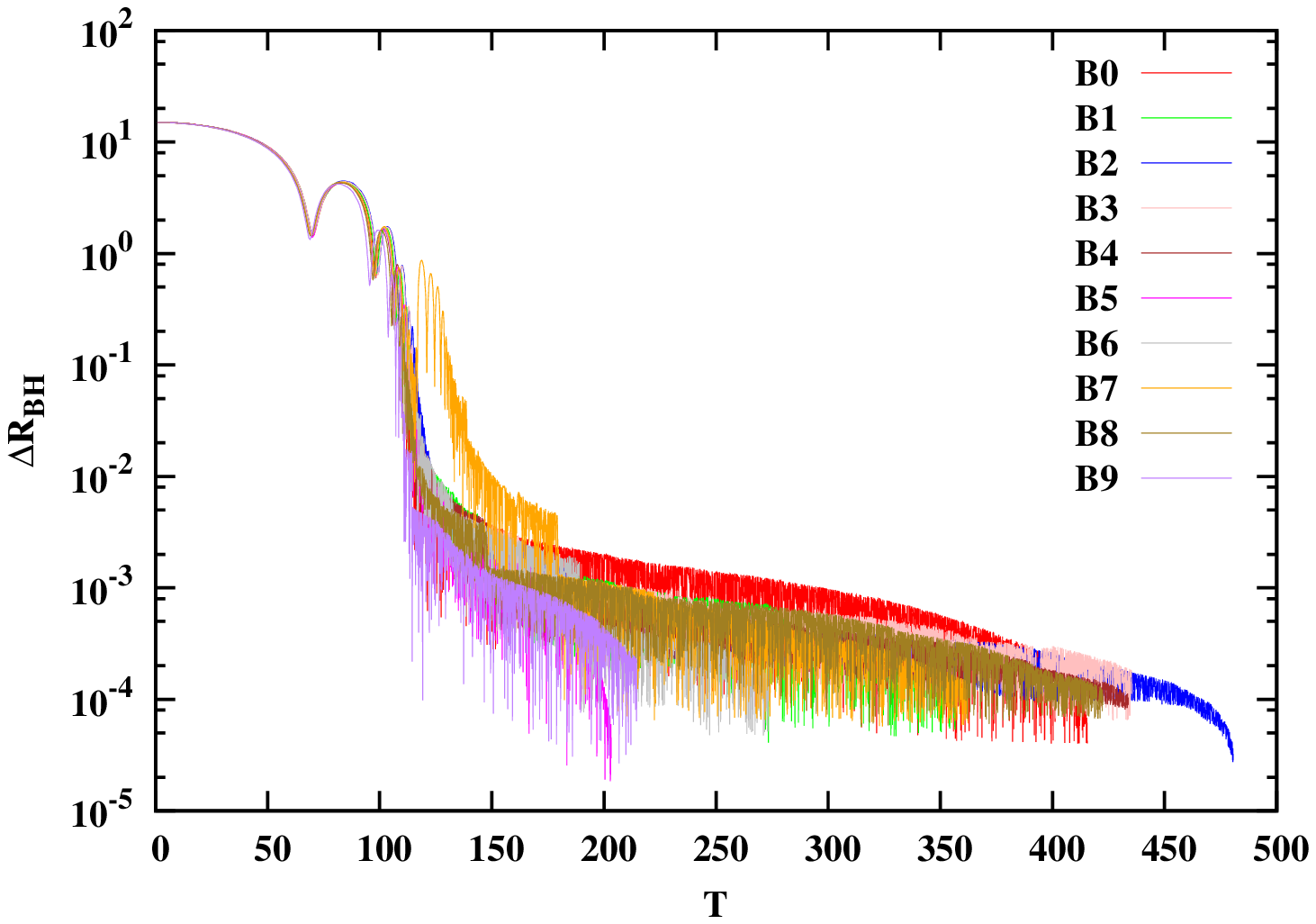}}
  }
  \centerline{
  \resizebox{0.9\hsize}{!}{\includegraphics[angle=270]{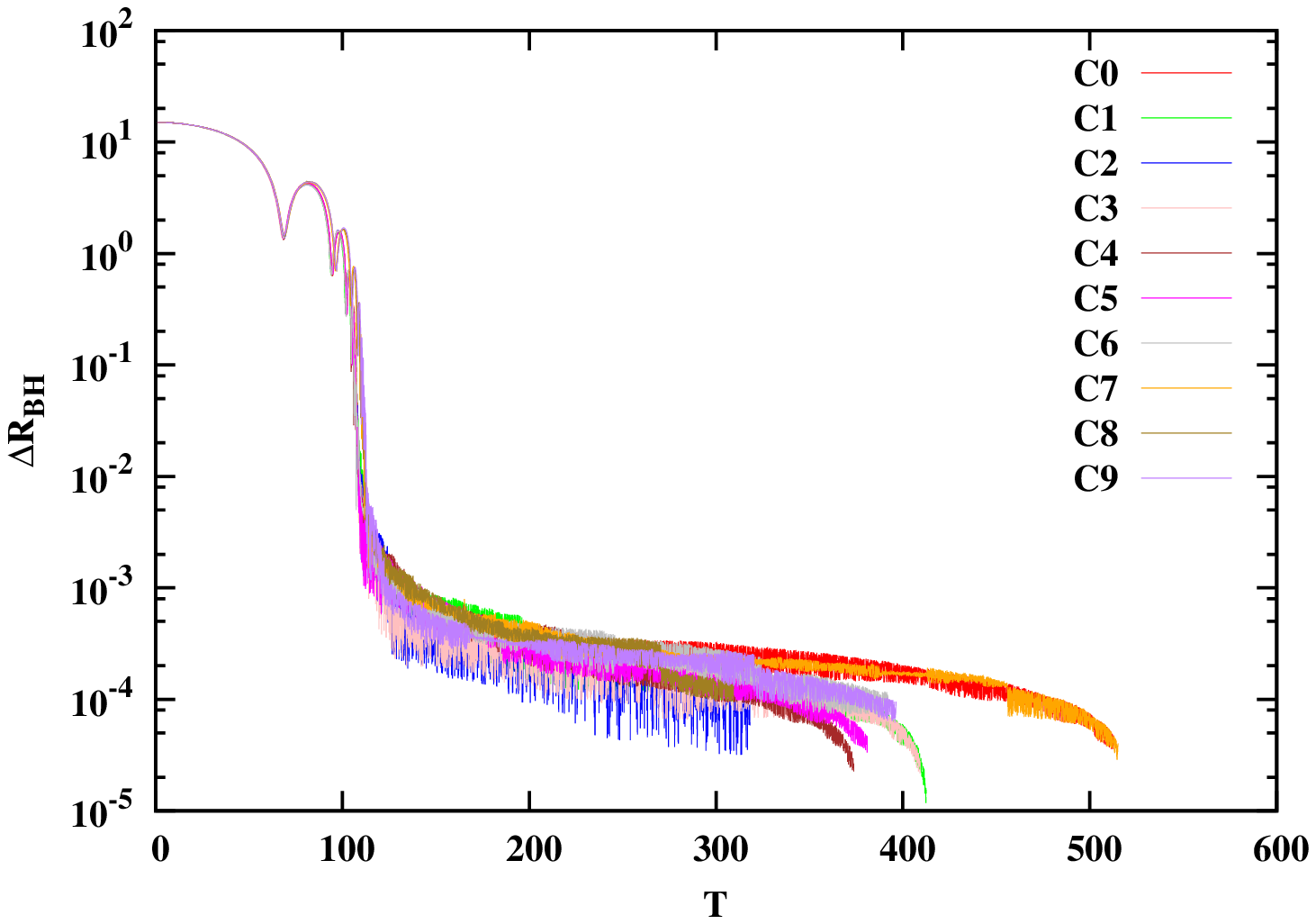}}
  }
\caption[]{
Shrinking of the separation between the SMBHs in galaxy mergers (top panel for models A, middle panel for models B and bottom panel for models C). Separation and time are in model units.
} \label{Fig1}
\end{figure}

We discuss first the hardening of the SMBH binaries. Figure \ref{Fig1} shows how the separation between the SMBHs shrink initially due to the galaxy merger, later on as a SMBH binary forms and hardens due to dynamical friction, in the hard binary phase due to three-body scattering and finally due to GW emission. As the galaxies merge, the separation between the
two black holes shrink below 1 (model) unit. Galaxies are merged at
almost the same time at $T  \sim 100$ after starting the simulations for all  models because of the same masses and orbits. Then we witness a fast decay in the
SMBH binary separation due to dynamical friction. Dynamical friction is more efficient
for steep cusps due to a higher central density

resulting in a faster orbital decay for case C compared to case A. However, dynamical friction becomes inefficient, when the orbital velocity of the binary is significantly larger than the velocity dispersion, i.e. when the binary gets hard. In case C this happens at a smaller separation due to the larger velocity dispersion.

We can see that for models A
dynamical friction becomes less efficient at a separation between the two SMBHs
of roughly 0.01 in model units, whereas for models C the same happened
at a 10 times smaller separation.

As a consequence of these two competing effects the transition to the 3-body scattering phase takes place at roughly the same time $T\sim 150$. The oscillations of the separation due to the eccentric SMBH binary orbit are not fully resolved in the plots due to the short orbital time compared to the larger output timesteps.

\begin{figure}
\centerline{
  \resizebox{0.9\hsize}{!}{\includegraphics[angle=270]{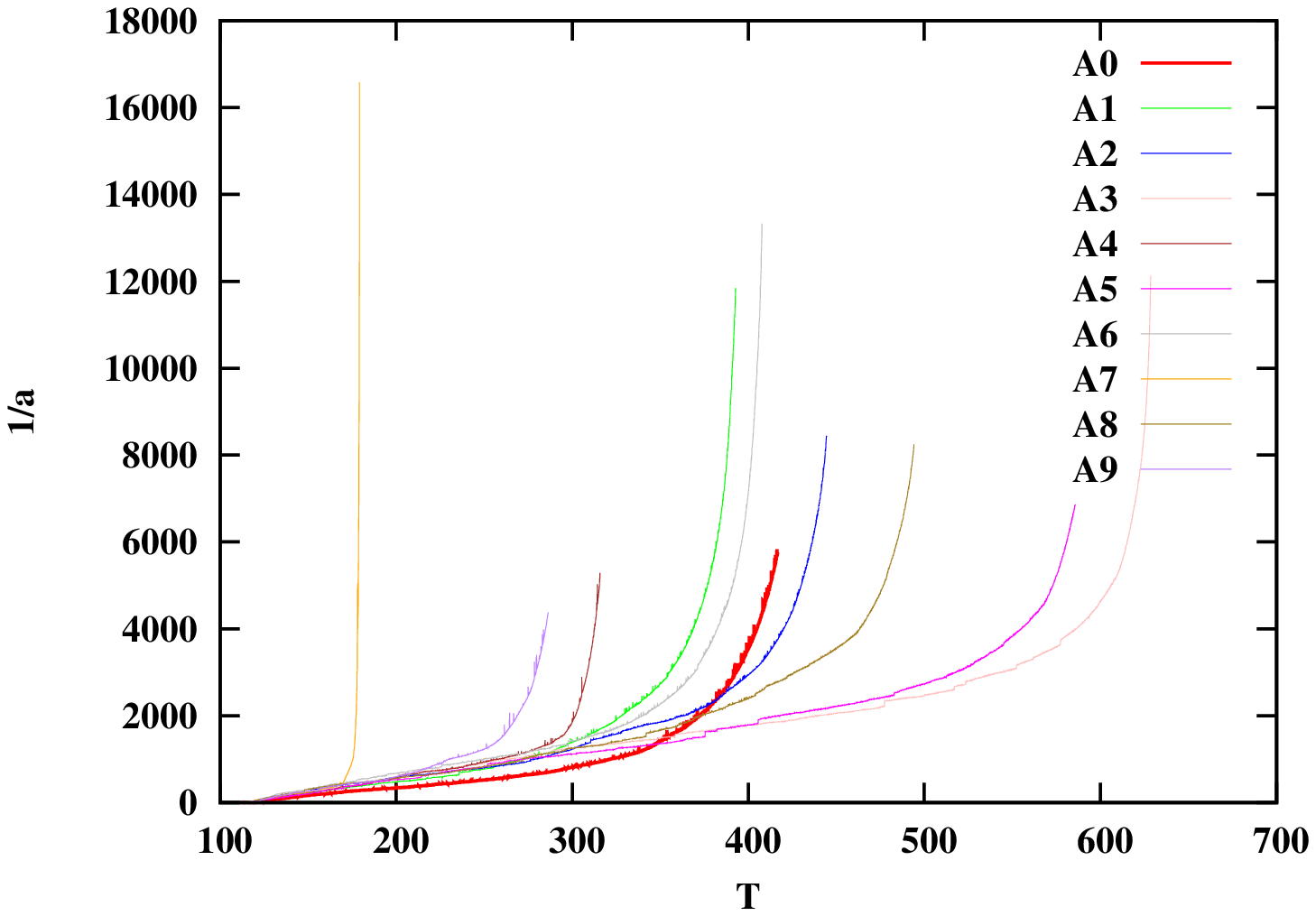}}
  }
\centerline{
  \resizebox{0.9\hsize}{!}{\includegraphics[angle=270]{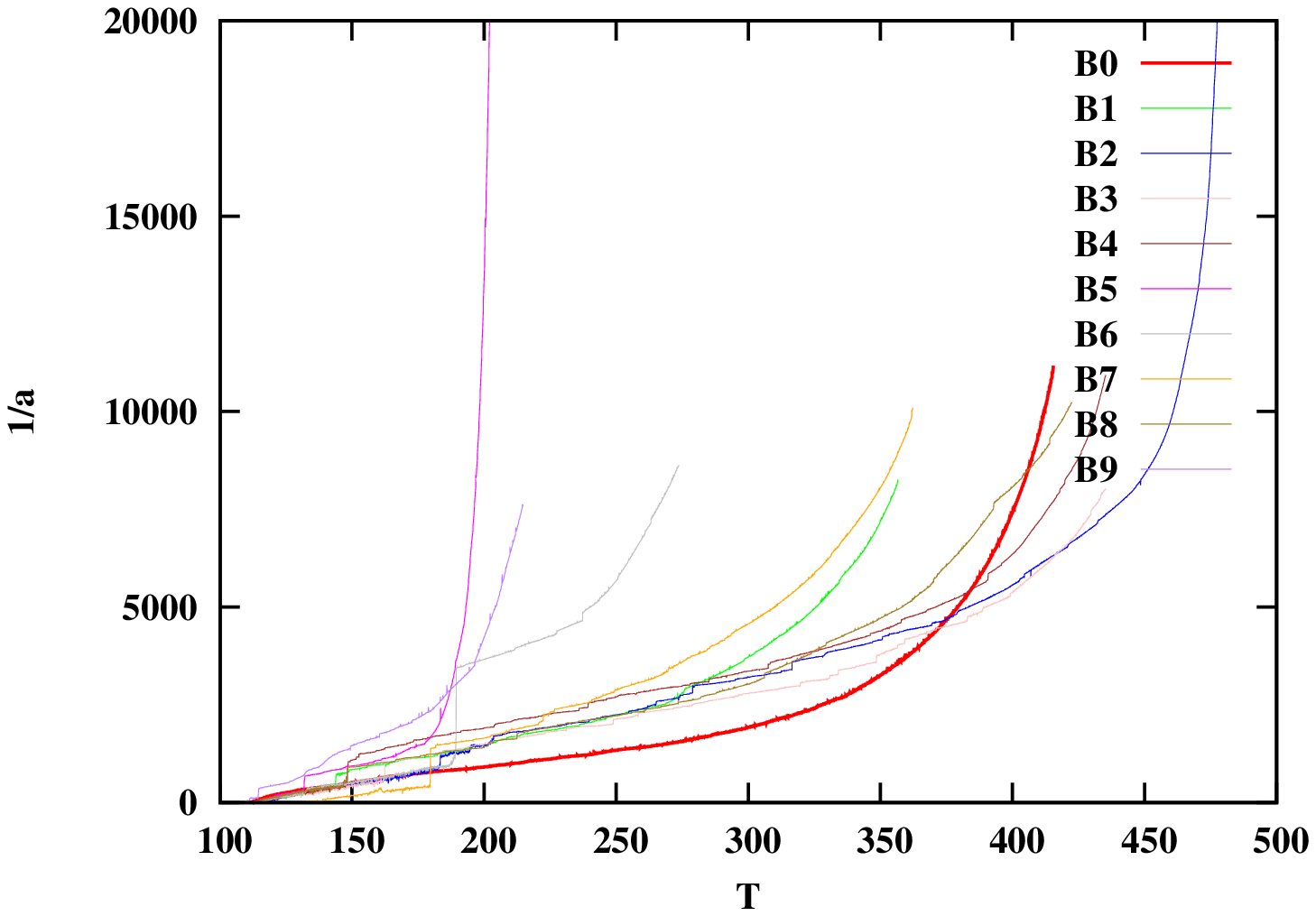}}
  }
  \centerline{
  \resizebox{0.9\hsize}{!}{\includegraphics[angle=270]{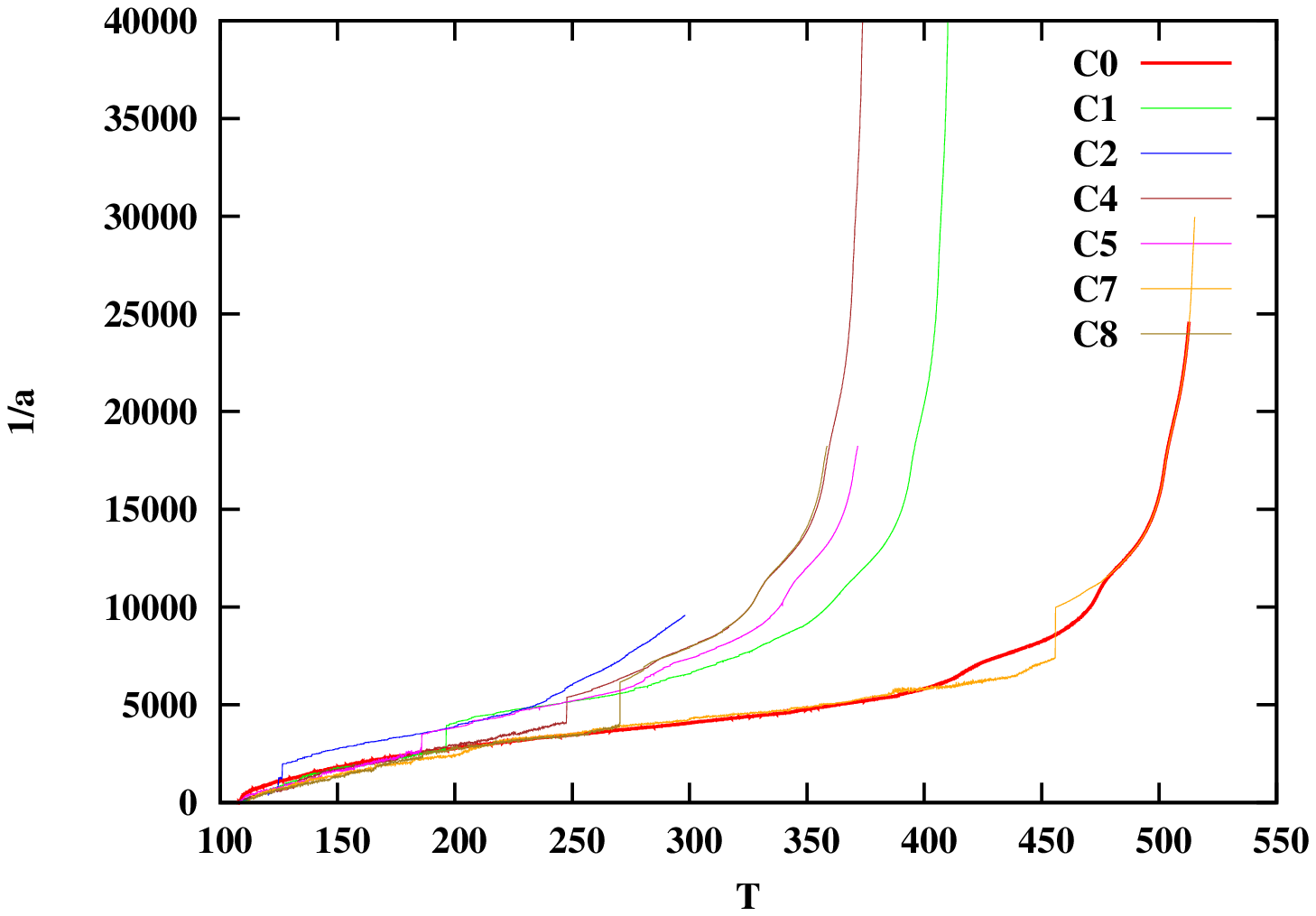}}
  }
\caption[]{
Evolution of the inverse semi-major axis (top panel for models A, middle panel for models B and bottom panel for models C). The single mass model 0 is represented with a thicker line in each plot.
} \label{Fig2}
\end{figure}

The evolution of the inverse semi-major axis of the SMBH binaries (which is a measure of the binding energy) for all our merger models is shown in figure \ref{Fig2}. We see that the inverse semi-major axis of the binaries evolve at a constant rate due to three body scattering. The final phase, where the energy loss by GWs dominates, is characterised by an accelerated hardening. The onset of this phase and the final merging times are consistent with the analytic estimates combining a constant hardening rate $s_\mathrm{3body}$ and the orbit averaged hardening rate from  \citet{pet63}.

We estimated the SMBH binary hardening rates  $s=\frac{d}{dT}(1/a)$ in the stellar dynamical hardening regime by fitting straight lines to $a^{-1}(t)$ (see table \ref{TableB}). The hardening rates are systematically higher for steeper cusps with higher values of $\gamma$ as already shown in \cite{khan12}.

The last rows in table \ref{TableB} shows the mean hardening rate $\langle s \rangle$ and the rms scatter of the simulations with SMF for each series. The hardening rates are approximately $30-40 \%$ higher for SMBH binaries evolving in merger remnants formed as a result of mergers of galaxies having a SMF compared to the
single-mass simulations A0, B0, C0 without SMF.

\begin{table*}
\caption{Galaxy merger runs} 
\centering
\begin{tabular}{c c c c | c c c c | c c c c c}
\hline
Run & $s$ & $e$ & T$_{coal}$ (Gyr) & Run & $s$ & $e$ & T$_{coal}$ (Gyr) & Run & $s$ & $e$ & T$_{coal}$ (Gyr)\\
\hline
A0 & 4.36 & 0.87 & 1.16 & B0 &  9.07 & 0.85 & 0.43 & C0 & 14.8  & 0.15 & 0.38\\
A1 & 6.35 & 0.82 & 1.10 & B1 & 12.41 & 0.72 & 0.38 & C1 & 19.88 & 0.41 & 0.30\\
A2 & 5.22 & 0.84 & 1.26 & B2 & 12.03 & 0.56 & 0.48 & C2 & 20.04 & 0.58 & 0.26\\
A3 & 5.99 & 0.56 & 1.78 & B3 & 12.98 & 0.60 & 0.47 & C3 & 19.71 & 0.36 & 0.30\\
A4 & 5.97 & 0.84 & 1.40 & B4 & 12.72 & 0.46 & 0.47 & C4 & 21.45 & 0.34 & 0.26\\
A5 & 6.38 & 0.53 & 1.70 & B5 & 13.20 & 0.87 & 0.21 & C5 & 20.18 & 0.28 & 0.27\\
A6 & 7.22 & 0.85 & 1.25 & B6 & 13.24 & 0.77 & 0.31 & C6 & 21.12 & 0.37 & 0.34\\
A7 & 7.42 & 0.93 & 0.56 & B7 & 11.94 & 0.82 & 0.39 & C7 & 17.64 & 0.19 & 0.38\\
A8 & 5.68 & 0.76 & 1.42 & B8 & 12.21 & 0.65 & 0.47 & C8 & 20.28 & 0.24 & 0.26\\
A9 & 5.34 & 0.86 & 0.88 & B9 & 12.43 & 0.85 & 0.25 & C9 & 20.81 & 0.29 & 0.33\\
\hline
mean & 6.17  & 0.78 & 1.26 & & 12.57  & 0.70 & 0.38 & & 20.12  & 0.34 & 0.30 \\
scatter & 0.72 & 0.13 & 0.36& & 0.46 & 0.13 & 0.10 & & 1.03 & 0.1 & 0.04\\
\end{tabular}\label{TableB}
\\
\tablefoot{Columns 1,5,9: Merger runs for galaxies having $\gamma=0.5,1.0, \& 1.5$, respectively.  Columns 2,6,10: SMBH hardening rates in the stellar dynamical regime. Columns 3,7,11: SMBH binary average eccentricity in the stellar dynamical regime. Columns 4,8,12: SMBH coalescence times.

The two lines at the bottom of the table show the mean values and root mean square scatter for the SMF simulations for each series.
}
\end{table*}

\begin{figure}
\centerline{
  \resizebox{0.9\hsize}{!}{\includegraphics[angle=270]{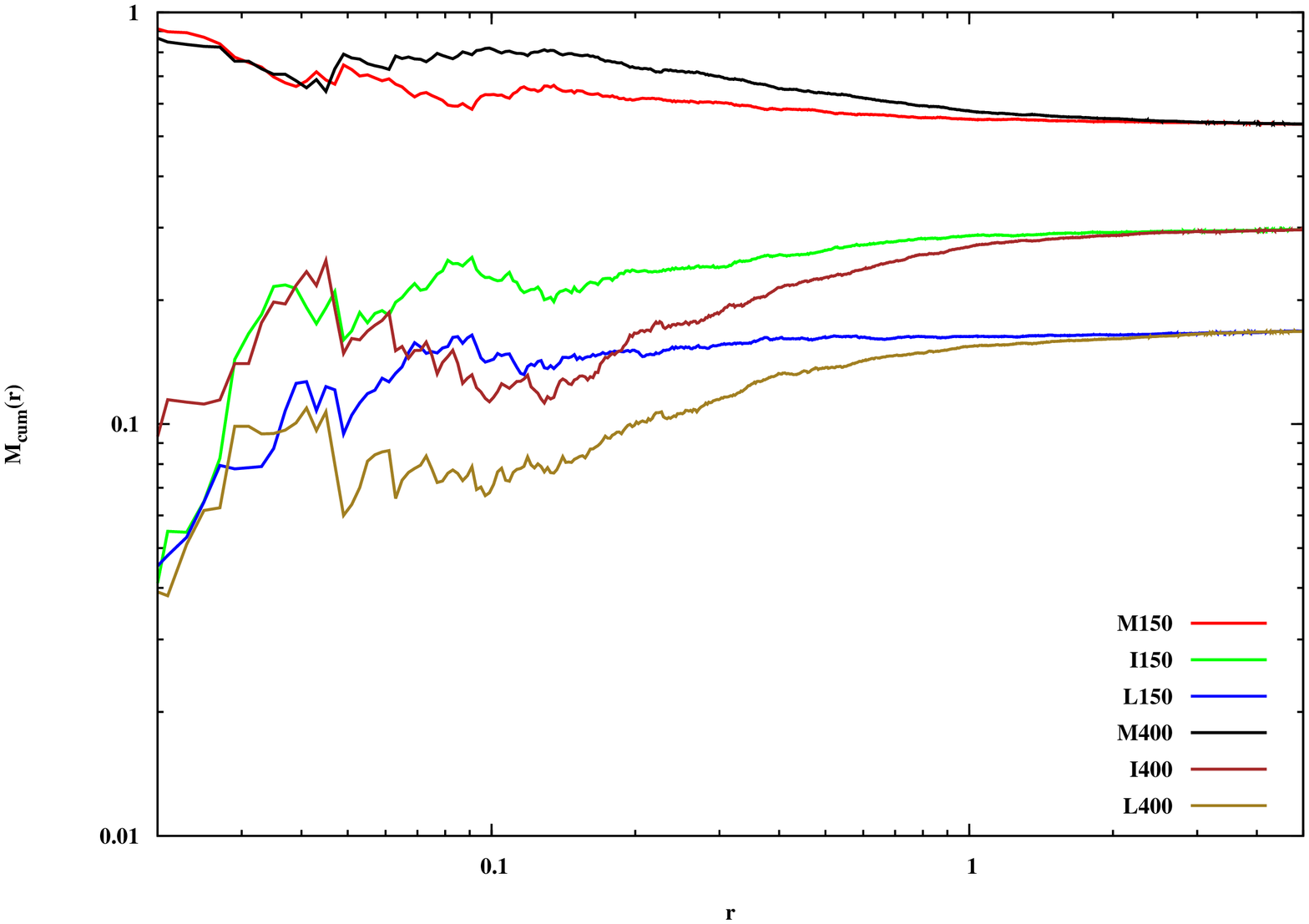}}
  }
\centerline{
  \resizebox{0.9\hsize}{!}{\includegraphics[angle=270]{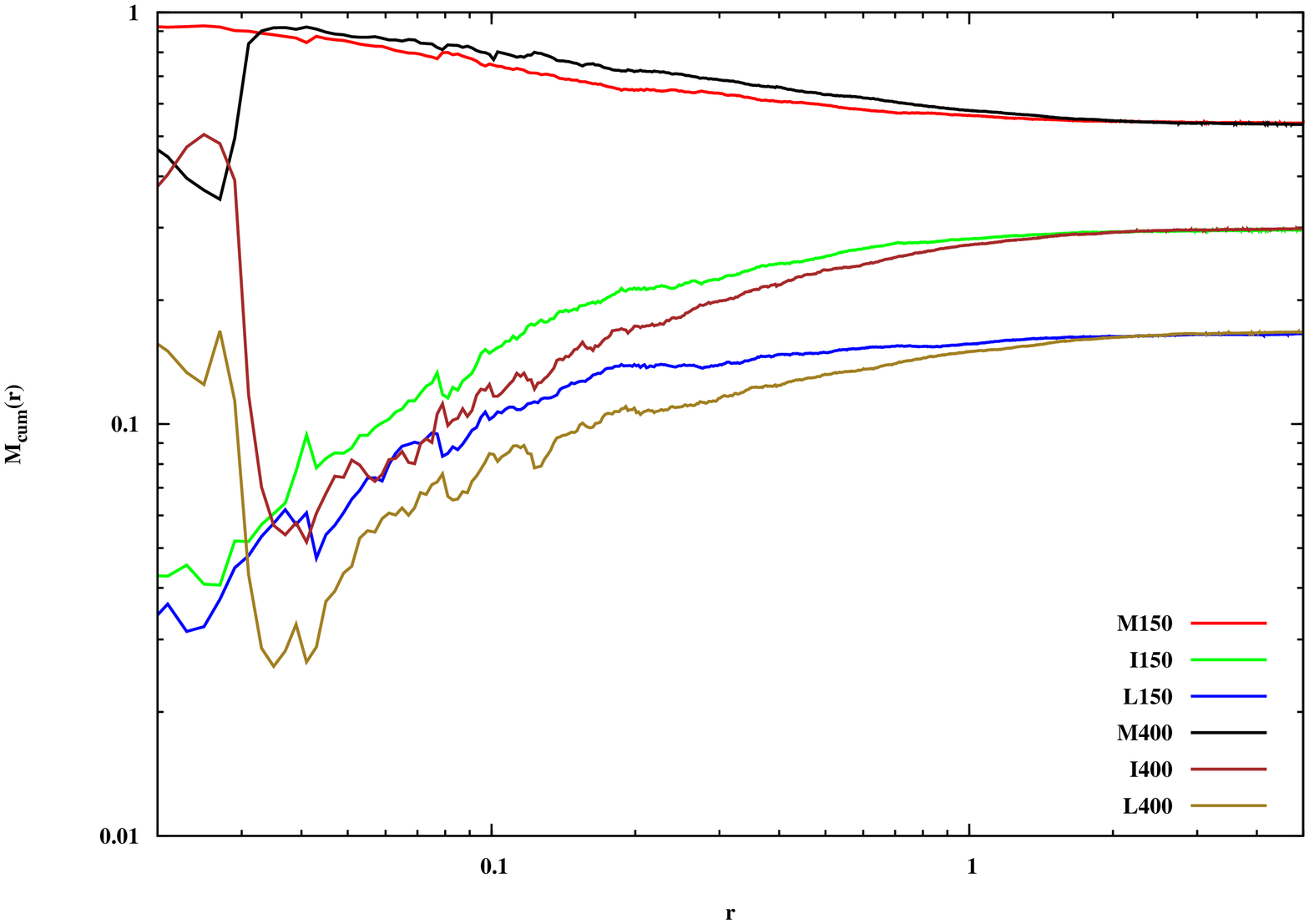}}
  }
  \centerline{
  \resizebox{0.9\hsize}{!}{\includegraphics[angle=270]{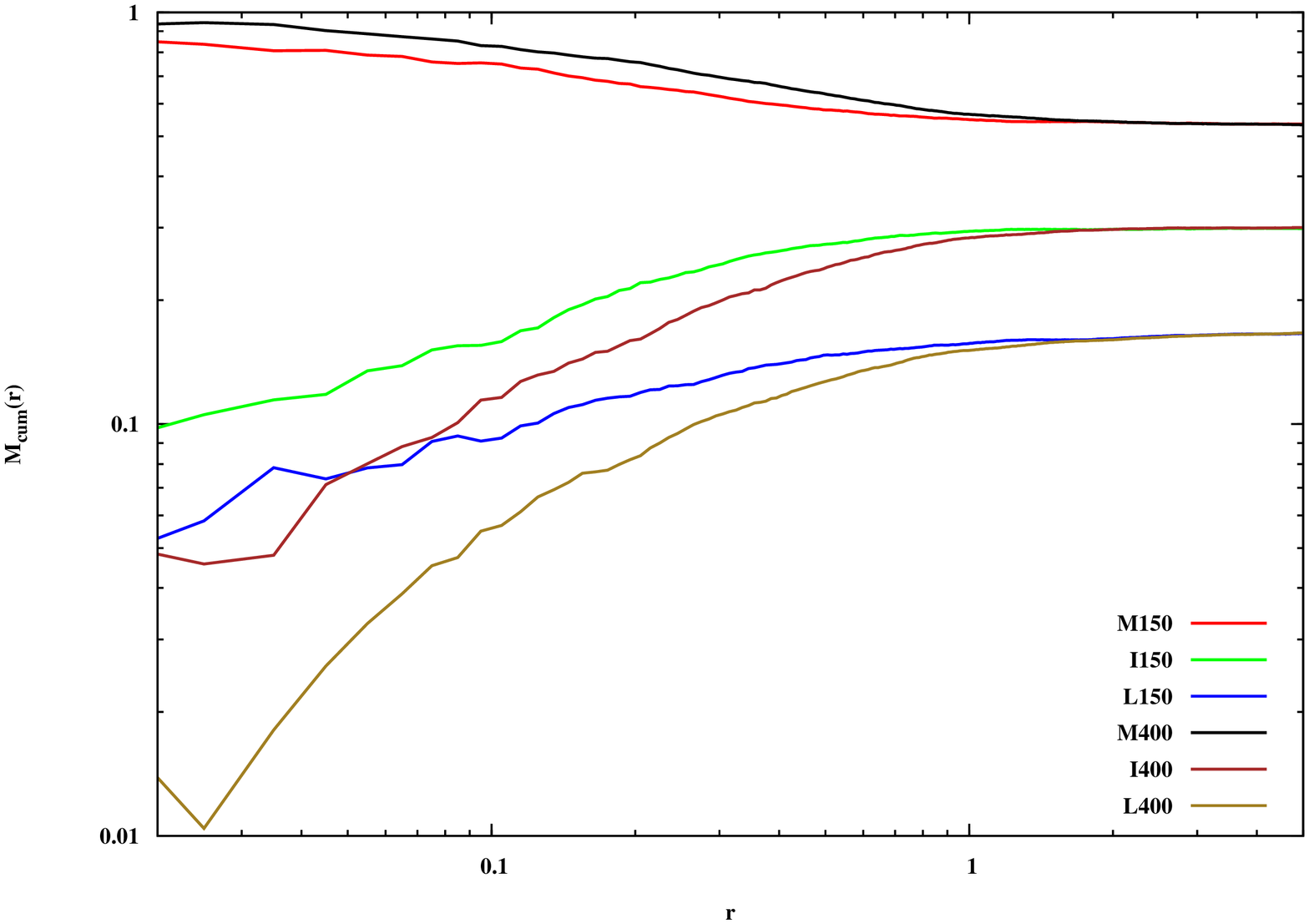}}
  }
\caption[]{
Fractional contribution of each stellar mass bin (M, I, L for massive, intermediate and low mass, resp.) to the cumulative mass profiles at an early ($T=150$) and late ($T=400$) evolution time for representative models A1 (top), B2 (middle) and C1 (bottom).
} \label{Fig3}
\end{figure}

If the stellar population is well mixed in phase space, there is no impact of a SMF on the hardening rate expected. 
The hard binary phase, where three body encounters dominate the energy extraction from the SMBH binary, corresponds to the low velocity limit for the intruders in three body scattering. For low-mass perturbers (mass ratios below 1:10 with respect to the secondary SMBH) the mean energy loss of the SMBH binary is proportional to the intruder mass \citep[see e.g.]{hills80,qui96,ses10}. For larger mass ratios, the energy loss is sublinear leading to a reduced hardening rate. As a consequence the hardening rate should depend only on the mass density distribution and the kinematic properties and scales as
\begin{equation}
s = \frac{G\rho_\mathrm{f}}{\sigma_\mathrm{f}} \,H, \label{sH}
\end{equation}
with the dimensionless hardening parameter $H$ \citep{ses15}. Density and velocity dispersion $\rho_\mathrm{f},\sigma_\mathrm{f}$ are to be taken at the radius $r_\mathrm{f}$, usually the influence radius of the binary \citep[e.g.]{ses10}.
In a first test to find the reason for the enhanced hardening rate, we compare the mass density and velocity dispersion profiles as well as the anisotropy profiles of the single-mass and the SMF simulations and do not find any differences above the noise level. But the number density profiles are different, which shows that the mean particle mass is increasing with decreasing distance from the SMBHs.
In order to see whether or not these increased hardening rates witnessed in SMF runs are due to mass segregation, we investigated the mass profiles of particle species of different mass ranges. We choose three mass species of particles (in numbers), most massive $12\%$ (M), intermediate $30\%$ (I) and least massive $58\%$ (L). For figure \ref{Fig3} we selected representative runs  (A1, B2, C1) of each $\gamma$ in our merger simulations. The plots show the fractional contribution of each stellar mass bin to the cumulative mass profile for two different times. We can clearly see that the three species have very different mass profile showing mass segregation for the most massive species (the estimated mass segregation time scale is $\sim 100$ time units). The steeper density profile of the high-mass stars goes hand-in-hand with a smaller velocity dispersion and vice versa for the low-mass stars. If we apply Eq. \ref{sH} to each mass component separately and add-up the contributions to total hardening rate (adopting a universal H), we find a slightly larger value for s, since the ratio $\rho_\mathrm{f}/\sigma_\mathrm{f}$ is smaller than the sum $\rho_\mathrm{i}/\sigma_\mathrm{i}$. But this effect is very small and does not explain the enhanced hardening rate for the SMF case.

\begin{figure}
\centerline{
  \resizebox{0.9\hsize}{!}{\includegraphics[angle=270]{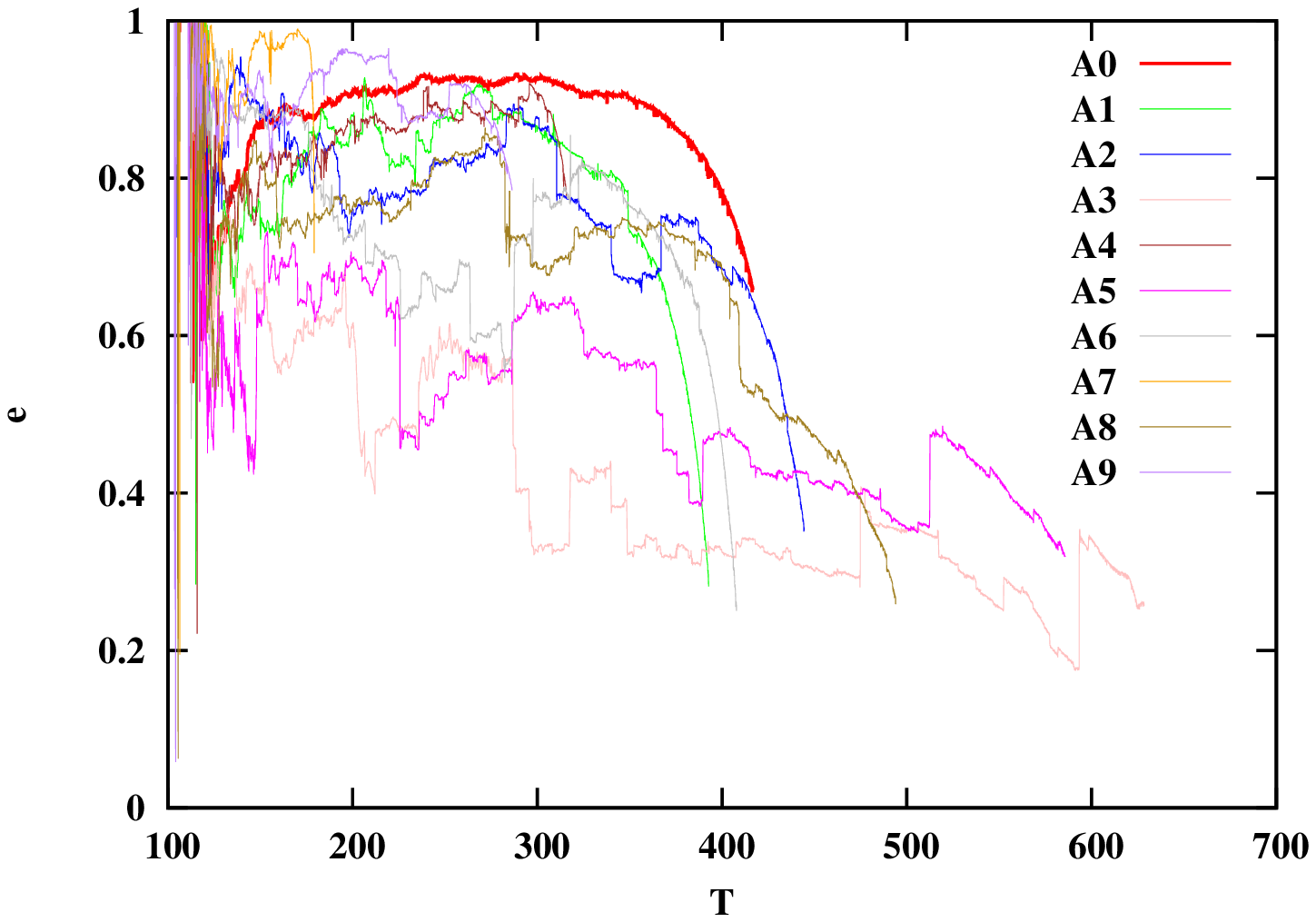}}
  }
\centerline{
  \resizebox{0.9\hsize}{!}{\includegraphics[angle=270]{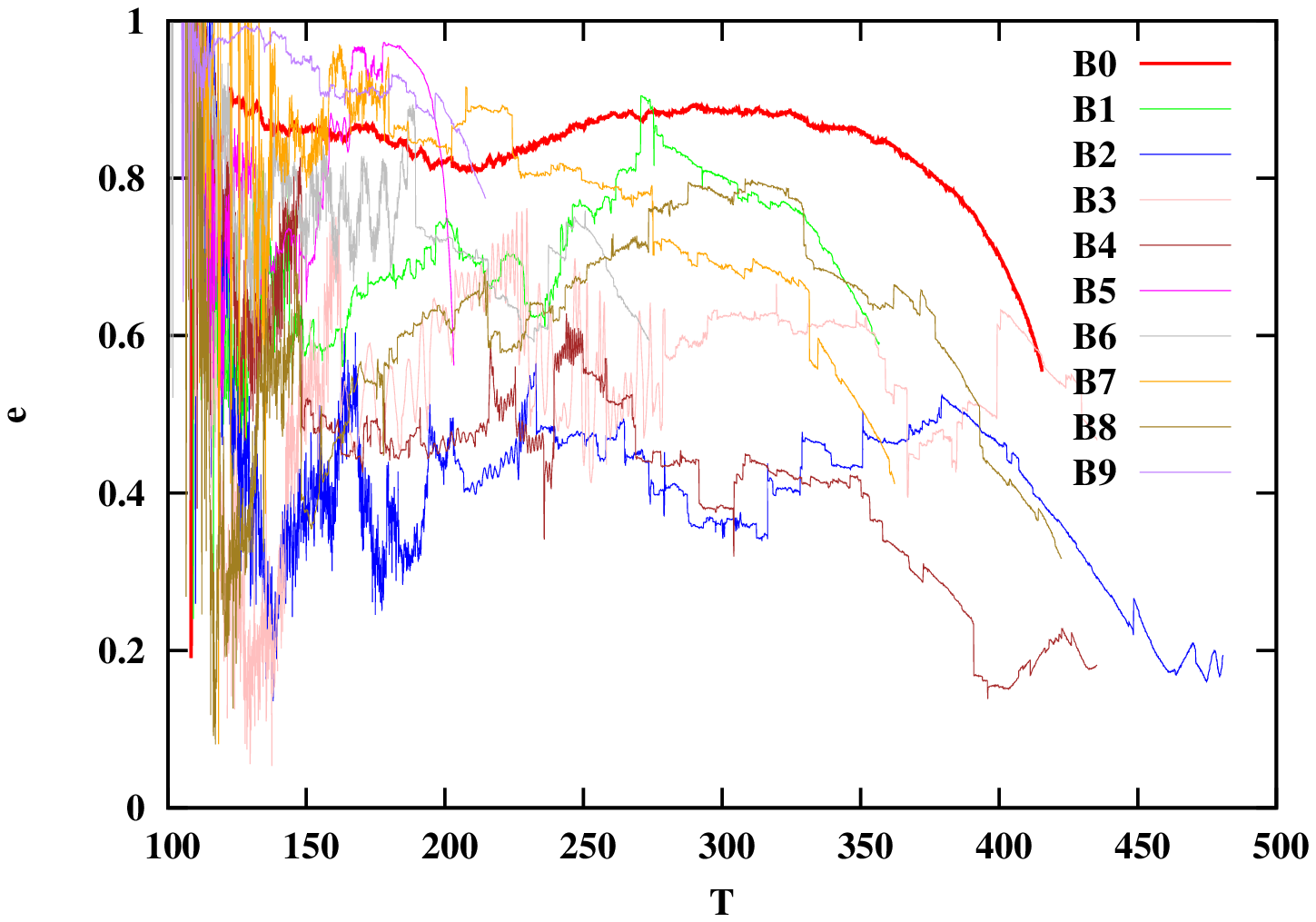}}
  }
  \centerline{
  \resizebox{0.9\hsize}{!}{\includegraphics[angle=270]{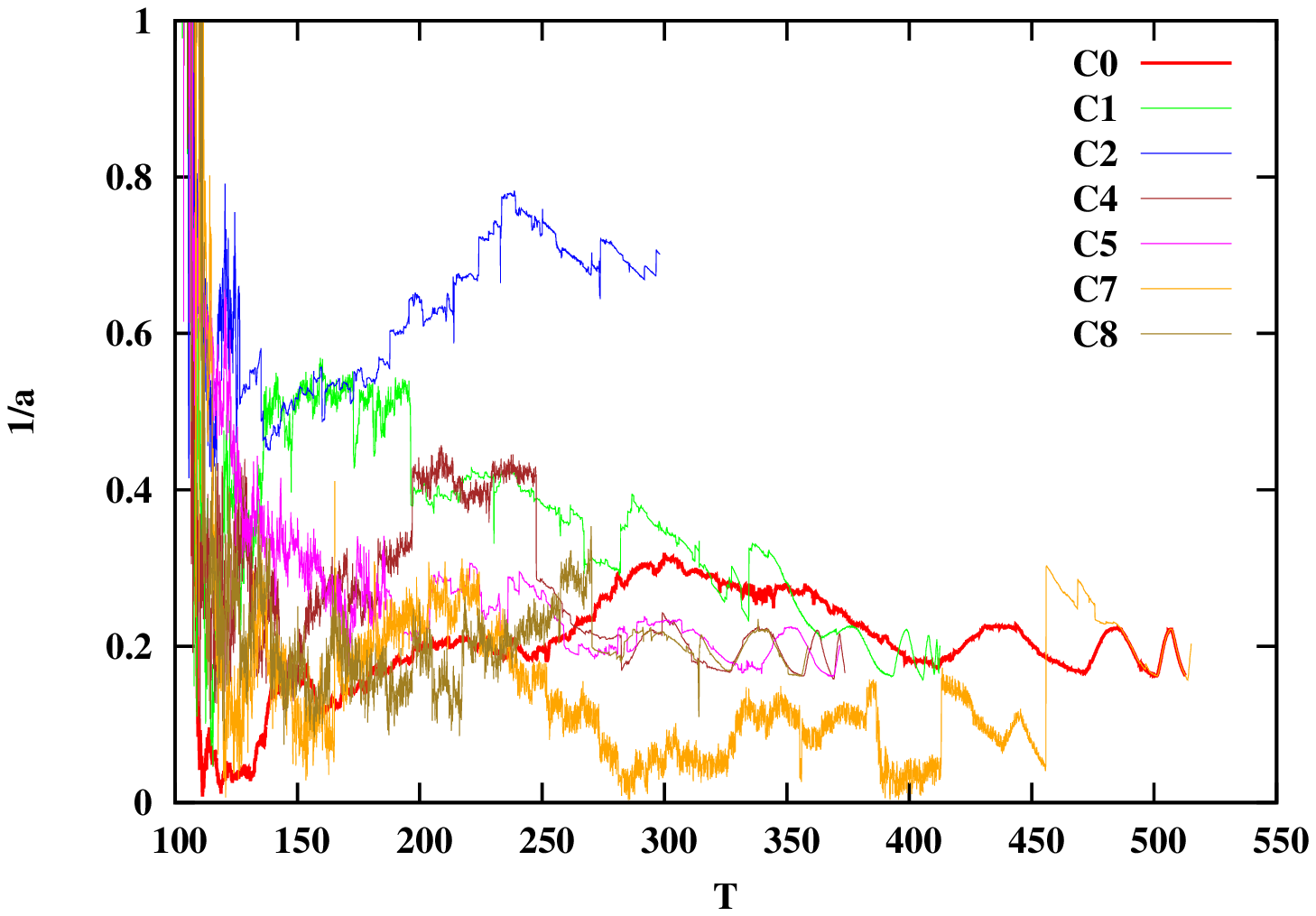}}
  }
\caption[]{
Evolution of SMBH binary eccentricities (top panel for models A, middle panel for models B and bottom panel for models C). Model 0 is represented with thicker line.
} \label{Fig4}
\end{figure}

It is well known that the quantification of the eccentricity evolution, which is connected to the angular momentum of the binary \citep[e.g.]{mik92,qui96,ses10}, is much harder than for the binding energy evolution. The main reason is that angular momentum changes are first order perturbations leading to a high sensitivity of the eccentricity on the random properties of the individual encounters. 
The evolution of the SMBH binary eccentricities $e$ is shown in figure \ref{Fig4}. In table  \ref{TableB} the mean eccentricity for each run is listed as well as averages over the nine SMF runs for each series. We observe strong fluctuations of the individual eccentricities, which are significantly larger for the SMF runs. A detailed discussion is given in Sect. \ref{N-test}. We notice that the mean eccentricities of the SMF runs show a large scatter in each series. The eccentricities of the single-mass runs A0 and B0 are on the high-eccentricity end, whereas for the steep cusp C0, the eccentricity is on the low end. We do not observe a correlation of the hardening rate with the eccentricity, but at least for the shallow and intermediate cusp series A and B the coalescence times depend strongly on $(1-e^2)$ as expected from the inset of energy loss by GW emission. For the low eccentricities of case C, the impact of variations in $e$ are small.

\begin{figure}
\centerline{
  \resizebox{0.9\hsize}{!}{\includegraphics[angle=270]{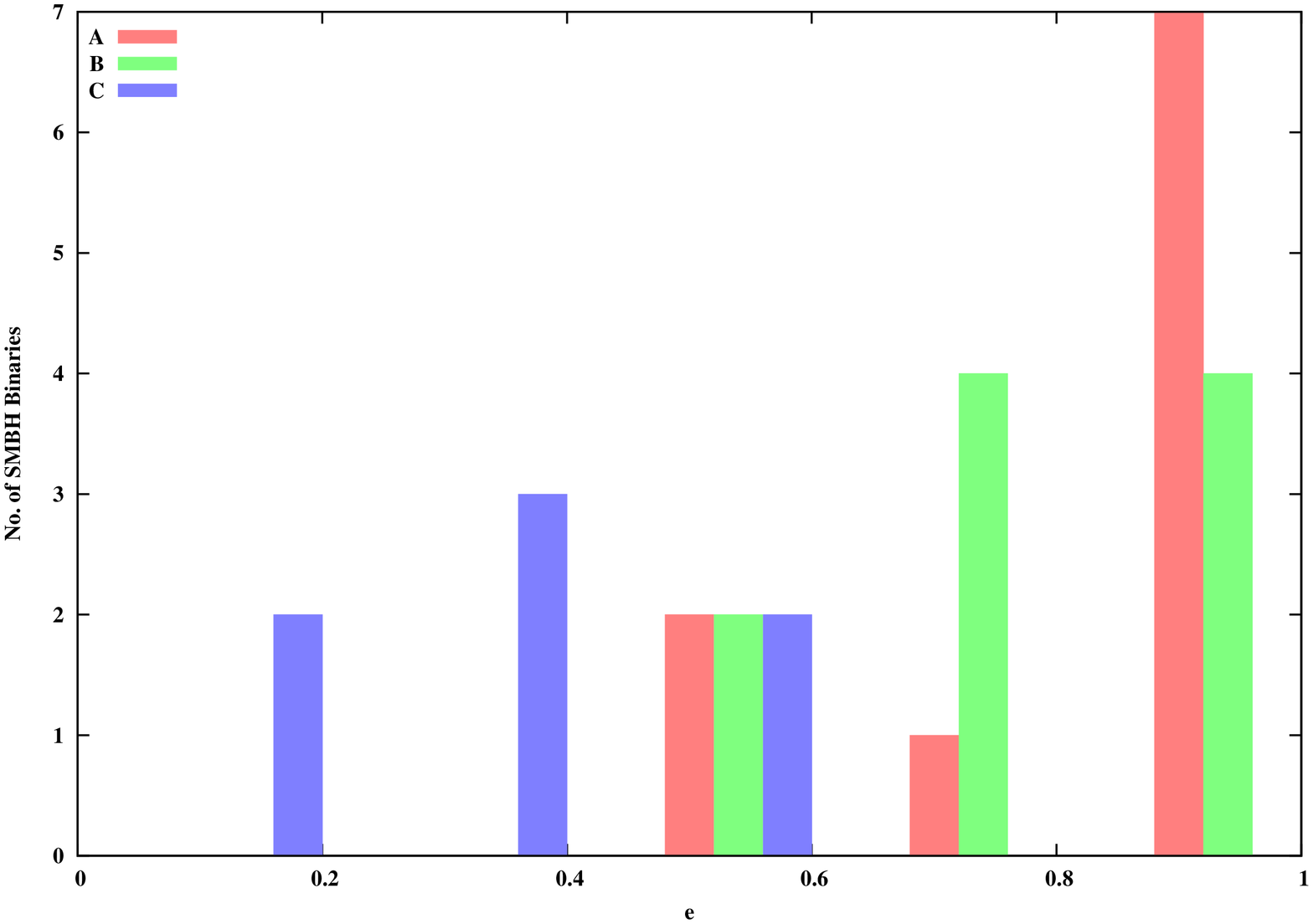}}
  }
\caption[]{
Histograms showing the number of SMBH binaries in eccentricity bins with width 0.2.
} \label{Fig5}
\end{figure}

We have plotted histograms of the average eccentricities in bins of 0.2  for all models in figure \ref{Fig5}. For models A, most binaries (7 out of 10) have a very high average eccentricity (in the range 0.8-1.0). Whereas for models B, this number becomes 4 for both 0.8 -1.0 and 0.6 - 0.8 bins. For both series A and B, the number of SMBH binaries in the eccentricity range 0.4 - 0.6 is 2, whereas there is no binary having an eccentricity below 0.4. The situation is very different for runs C, where the merging galaxies have steep inner profiles (dense nuclei) with $\gamma = 1.5$. The number of SMBH binaries peaks (5 out of 10) in the bin 0.2-0.4. There are three binaries in the bin 0 - 0.2 and two in the bin 0.4 - 0.6. This systematic trend to more circular binary orbits for steeper cusps is in accordance to the more effective circularisation by dynamical friction in steeper density profiles.


The coalescence times $T_{\mathrm{coal}}$ for SMBH binaries are also collected in table \ref{TableB}. 

The average coalescence time of SMBH binaries in the SMF runs A is 1.25\,Gyr. For SMBH binaries in runs B and C it is 0.39 and 0.31\,Gyr, respectively. 
In series A and B the rms variation of $\sim25\%$ is similar, whereas in series C the scatter is much smaller. 
The main reason for this variation is the scatter in eccentricity in the high eccentricity regime.

\begin{figure}
\centerline{
  \resizebox{0.9\hsize}{!}{\includegraphics[angle=270]{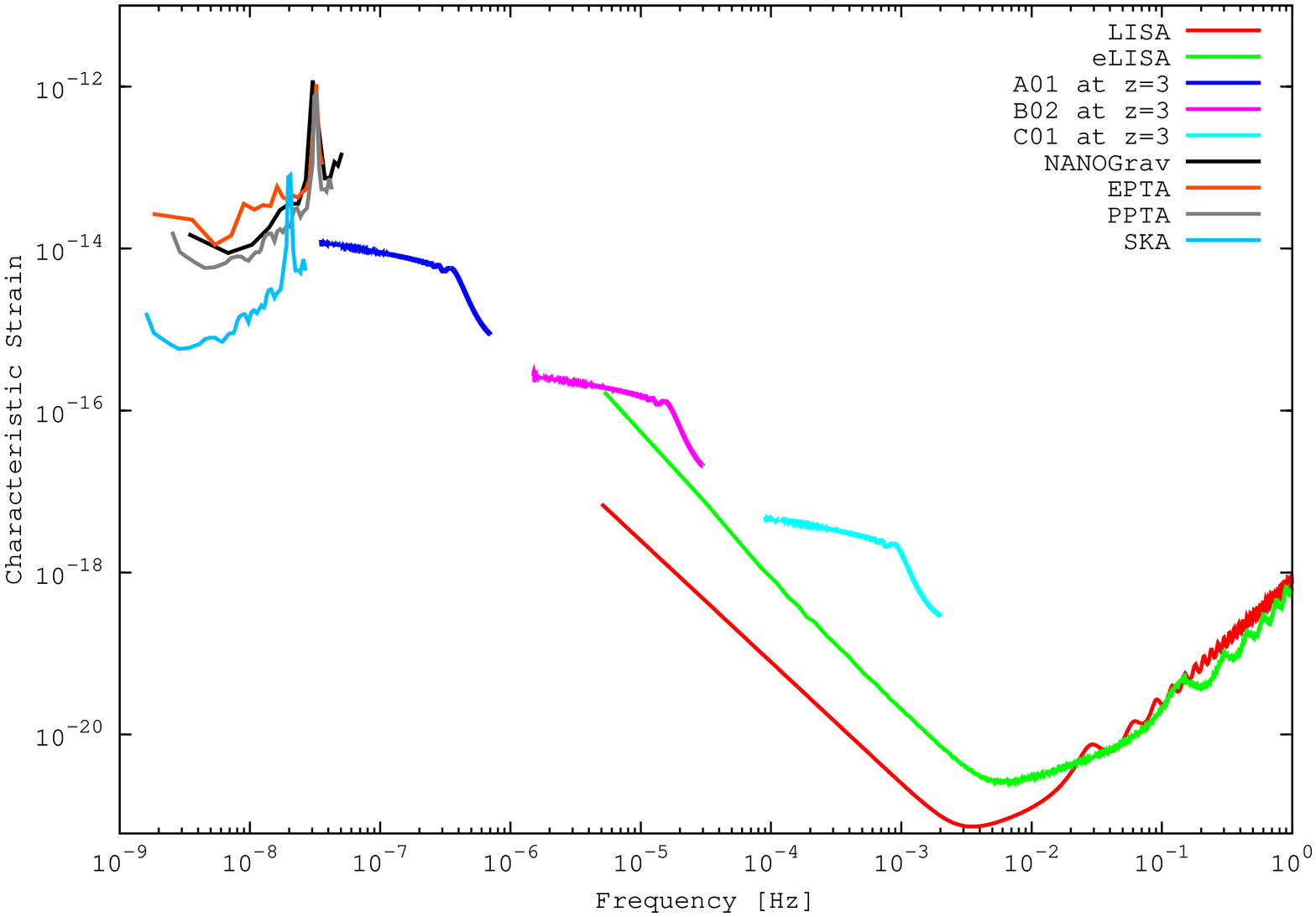}}
  }
\caption[]{
Comparison of the predicted characteristic strain for models A1, B2, and C1 at redshift $z=3.0$ and the sensitivity curves for various GW detectors. 
} \label{Fig6}
\end{figure}

In the late phase of the evolution the SMBH binaries emit low frequency GWs, which may be observable with GW detectors like various pulsar timing array experiments \citep{des16,rea16,nan15,ver16} and planned space borne GW observatory eLISA or LISA  \citep{amaro13,goat16,lisa17}. In figure \ref{Fig6}, we have calculated for the three selected cases A1, B2, and C1 the characteristic strain adopting a redshift of $z=3.0$ with a corresponding luminosity distance of $D=26$\,Gpc. At least the low mass SMBHs (in steep cusps) should be visible at the low frequency end of eLISA and LISA. The high mass end (with shallow cusps) is close to the frequency range and sensitivity of current PTAs.

\subsection{Convergence Tests}\label{N-test}

We notice a few sudden jumps in the $1/a$ evolution, especially for steeper cusp mergers. Sometimes there is also a jump in eccentricity, but not always. Our analysis shows that occasionally the SMBH binary and a massive stellar particle at the high end of the stellar mass function form a three body bound system surviving for a relatively long time. In this phase the slingshot ejection of a 4th body is able to produce these large jumps in $1/a$. For the determination of the hardening rate we have corrected for these unrealistic jumps by measuring $s$ in time intervals with no significant jumps. In the eccentricity $e$ evolution there are more jumps in the shallow cusp case A. These jumps are not dominating the overall evolution. Most of the fluctuations occur on longer timescales covering many orbital times and many scattering events. For reducing the effect of these random scatter significantly much higher particle numbers may be necessary. In this section we test the $N$ dependence of the hardening rate and the fluctuations in the eccentricity.

\begin{figure}
\centerline{
  \resizebox{0.9\hsize}{!}{\includegraphics[angle=270]{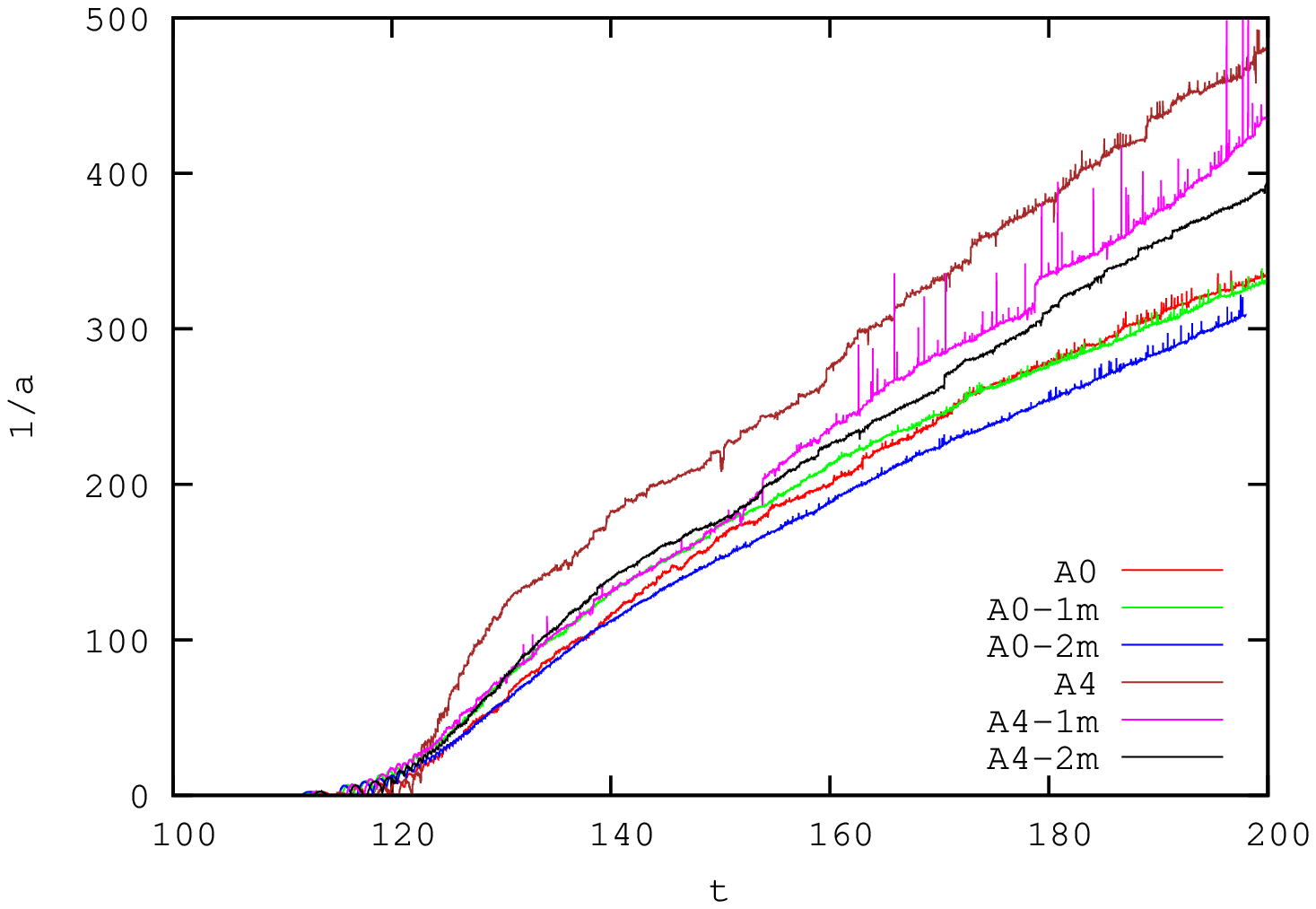}}
  }
\centerline{
  \resizebox{0.9\hsize}{!}{\includegraphics[angle=270]{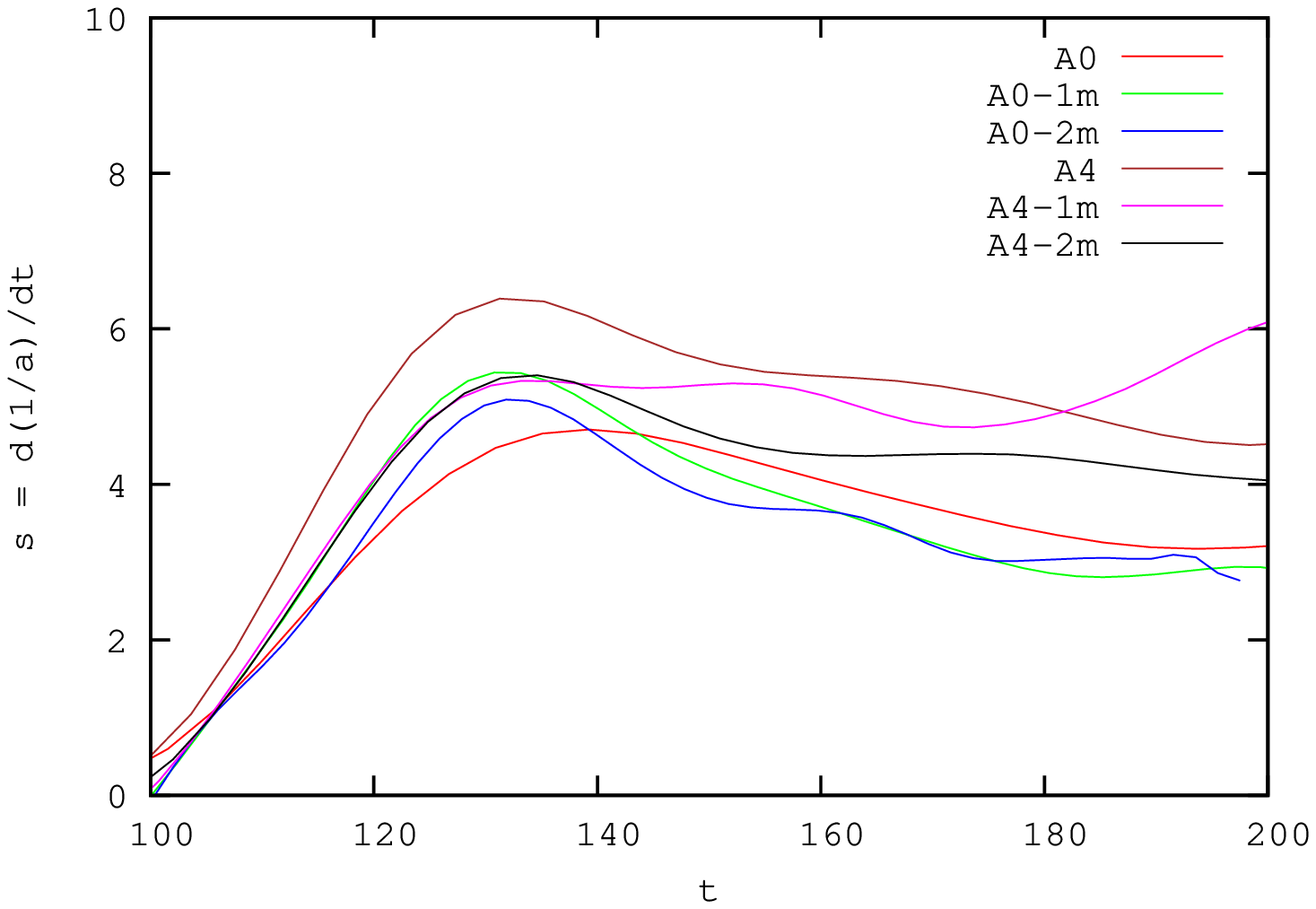}}
  }
  \centerline{
  \resizebox{0.9\hsize}{!}{\includegraphics[angle=270]{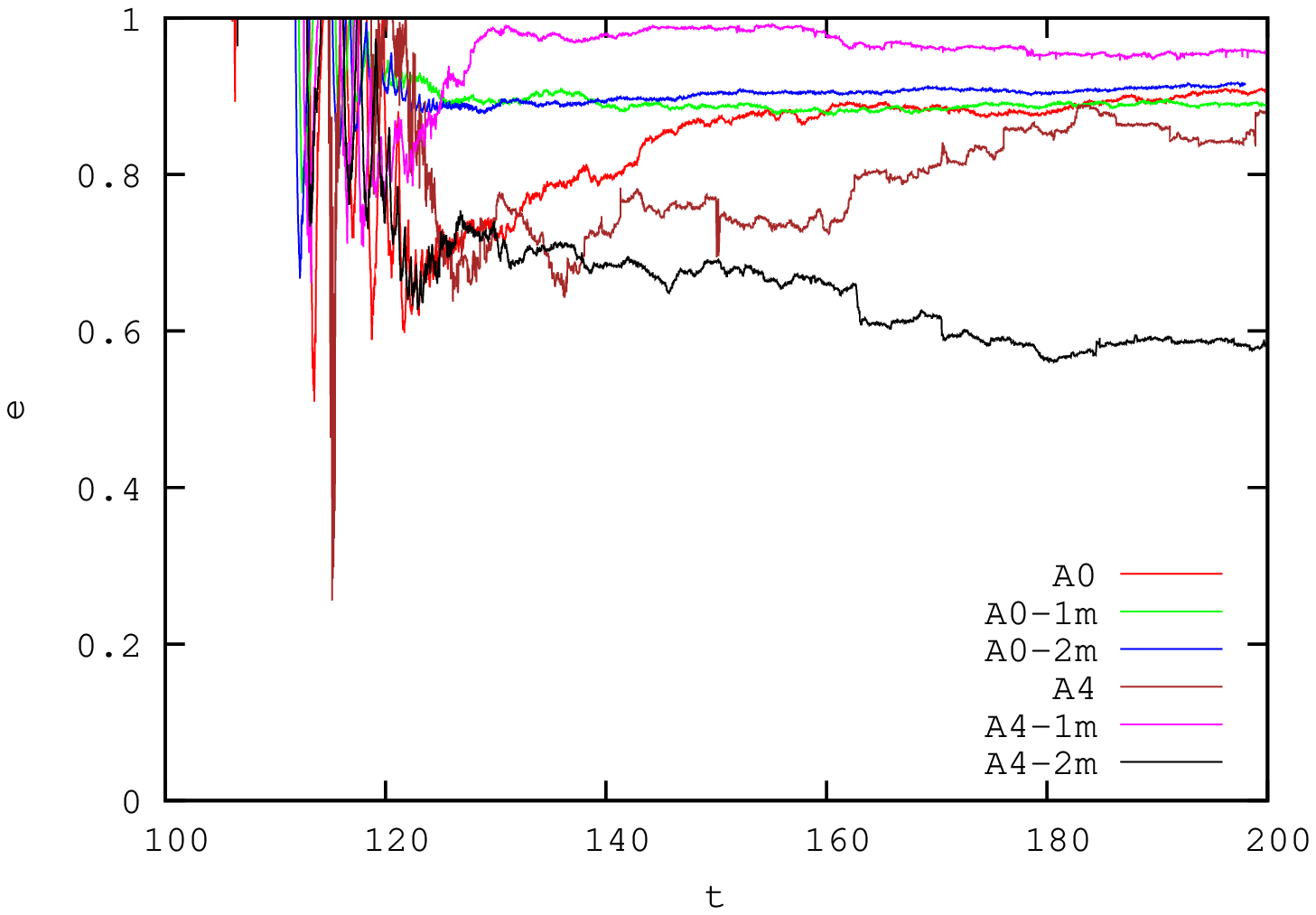}}
  }
\caption[]{
Evolution of SMBH binary inverse semi-major axis, hardening rates and eccentricity for our convergence test runs with $N$ = 400k, 1 and 2 million for the single-mass A0 and the SMF case A4.
} \label{Fig7}
\end{figure}

In order to test, how strong our results depend on the particle number $N$, we performed a convergence test for two of our runs where we use particles up to 2 million. We choose $\gamma = 0.5$ being a shallow profile, because it is expected that the shallow profile mergers take less computational time to complete. We choose two series of runs; A0, A0-1m and A0-2m are single-mass runs with 400k, 1 million and 2 million particles, whereas A4, A4-1m and A4-2m are SMF runs again with 400k, 1 million and 2 million particles. For our largest $N$ we have a five times smaller maximum mass ratio of the secondary SMBH and the stars of 1:77. The results of this study are presented in figure \ref{Fig7}. The top panel shows the inverse semimajor axis and the middle panel the corresponding smoothed hardening rates. In the early phase up to $T=150$, where dynamical friction is still active and the inner loss cone is not empty, there are differences in the evolution. Here we are interested in the stationary later phase, where the three body encounters dominate the hardening.

The N-independent evolution of $1/a$ witnessed by earlier studies for galaxy mergers is reproduced for the single-mass runs. The stationary phase of constant hardening rate is reached at about $T=200$. For the SMF runs a weak trend of decreasing $s$ with increasing $N$ may be there. However, the hardening rate of the 1 million run A4-1m show fluctuations breaking this trend at $T=200$. On top of this possible trend, we also notice that $s$ is systematically higher for all SMF runs when compared with their counterpart single-mass runs. Since the two body relaxation time and thus the mass segregation timescale depends on $N$, we expect a dependence (delay) of mass segregation and enhanced hardening with increasing $N$.
For the SMF cases we still see a considerable mass segregation of the high mass end as in the 400k case, but starting a bit later at $T=150$ consistent with the enhanced hardening rate. 
For the 1 and 2 million runs we observe the same density, velocity dispersion and anisotropy profile as for the 400k case. 
There is no significant difference of these global properties in phase space for the SMF and n-SMF cases. We calculate dimensionless hardening parameter H using equation \ref{sH} for our 2 million particle runs. H has a value of 14.82 for run A4-2m and a somewhat lower value of 12.2 for the A0-2m run, which reflects a similar difference in hardening rates in the two runs. We conclude that mass segregation has a more subtle effect on the distribution functions enhancing the encounter statistics in order to explain the higher hardening rates.

Another important parameter that enters the merger times is the SMBH binary eccentricity (bottom panel of figure \ref{Fig7}). For all single-mass runs there is a remarkable consistency in SMBH binary eccentricities of $\sim 0.9$ in the late stationary phase, despite the fact that for the 400k case A4 the eccentricity starts with a smaller value and increases as theoretically predicted. For the SMF runs we do not witness a strong match but eccentricities of all binaries are in the most favourable range of 0.6--1.0 for A runs and there is no systematic trend with increasing particle number. We note that the fluctuations in the eccentricity are not reduced significantly by increasing the particle number by a factor of five, although jumps are now much less prominent and the evolution of e is much smoother. This is a hint that they may arise form inhomogeneities in the distribution of particles leading to an anisotropic flux of interacting stars.
In order to reach the same maximum mass ratio of 1:500 as in the single-mass mergers, about 15 million particles would be needed, which is not feasible with the current computer facilities.

\section{Summary and Conclusion}\label{sum}

We have performed a statistical set of mergers of galaxies with a mass ratio of $q=1/4$ with shallow, intermediate and steep central density profiles. Each galaxy contains a central supermassive black hole SMBH with 1\% of the galaxy mass and the particles were realised with a Salpeter like stellar mass function SMF. We have used independent random realisations for the initial phase space distribution and for the stellar masses. The dynamical evolution of the galaxy pairs was performed with a direct N-body code including general relativistic effects of the SMBHs using post-Newtonian corrections. All simulations were performed until the final coalescence of the SMBHs by gravitational wave (GW) emission.

The total coalescence time is dominated by the length of the 3-body encounter phase, where the SMBH hard binary loose energy by slingshot encounters with stellar particles.
However, the end of this phase depends strongly on the eccentricity of the SMBH binary by the onset of GW emission.
The hardening rate $s$, the mean eccentricity $e$ and as a consequence the coalescence time $T_\mathrm{coal}$ show a significant scatter due to the random realisation of both, the phase space distribution and the stellar mass function. 
The scatter and fluctuations in the eccentricity does not decrease significantly (for the tested shallow cusp case), when increasing the number of particles by a factor or five for the SMF case. In contrast, the single-mass case shows a remarkable similar eccentricity for all particle numbers $N$. 

On top of this scatter, we observe significantly higher hardening rates for the steeper profiles due to the larger central densities. The mean eccentricities are smaller for steeper profiles which compensates partly for the faster evolution of the SMBH binaries due to the delayed influence of GW emission. Compared to the single-mass systems, the hardening rates of the systems with SMF are larger by $\sim 30-40\%$, whereas there is no systematic effect on the eccentricity observed. 

The enhanced hardening rate for the SMF simulations due to mass segregation is also seen in the dimensionless hardening rate $H$, because the density and velocity dispersion profiles are similar to the non-SMF cases. The reason must be hidden in a more subtle difference in the phase space distributions. This requires a much more detailed analysis of the system, which must be postponed to a future investigation.

We admit that the impact of the stellar mass function and mass segregation maybe overestimated, since the relaxation time in the simulations is too short due to the small particle number compared to realistic systems. 
Nevertheless the tendency of speeding up the SMBH binary evolution is interesting, because there are other ways to form a mass segregated galactic nucleus, for example by an inhomogeneous mixture of stellar populations with different ages.

We have applied the simulations to three representative galaxies for the three different density profiles. We found coalescence times of $0.30\pm 0.04$\,Gyr for the Milky Way (MW) (representing a steep slope of $\gamma=1.5$), $0.38\pm 0.10$\,Gyr for M31 with intermediate slope $\gamma=1.0$, and $1.26\pm 0.36$\,Gyr for M87 with a shallow slope $\gamma=0.5$. In all cases the coalescence time is short compared to the Hubble time and the expected time between two mergers in the present day universe. At high redshifts, where galaxies are compact and hence possess denser central regions, SMBH merger times can be orders of magnitude smaller \citep{kh16}. 

We have also calculated the strength of the GWs emitted in the final phase of the SMBH binary evolution and have shown that
low mass and intermediate mass mergers ($M_\bullet\sim 4\times 10^6 - 2\times 10^8\,M_\sun$) are visible with eLISA at the low frequency end at redshift $z=3$ (corresponding to a luminosity distance of $D=26$\,Gpc, whereas high-mass mergers  ($M_\bullet\sim 6\times 10^9\,M_\sun$) would be close to the frequency and sensitivity limit of current PTAs at that redshift.

\begin{acknowledgements}
We are thankful to Alberto Sesana for providing sensitivity curves data for various PTAs. F.M.K. acknowledges
support by the Excellenzinitiative II 'Mobilit\"atsma\ss nahmen
im Rahmen internationaler Forschungskooperationen
2015-16' of Heidelberg University. FMK also acknowledges support by Higher Education Commission of Pakistan through NRPU grant 4159.
The special GPU accelerated supercomputer Laohu funded
by NAOC/CAS and through the 'Qianren' special foreign experts
program of China (Silk Road Project), has been used for the simulations.
We thank the anonymous referee for his valuable comments, which improved the paper significantly.

\end{acknowledgements}

\bibliographystyle{aa}
\bibliography{mybibfile}

\end{document}